# Analytic Evaluation of some 2-, 3- and 4- Electron Atomic Integrals Containing Exponentially Correlated Functions of $r_{ij}$

Bholanath Padhy*

Department of Physics, Khallikote College, Brahmapur-760001, Odisha

* Email: bholanath.padhy@gmail.com

## ABSTRACT

A simple method is outlined for analytic evaluation of the basic 2-electron atomic integral with integrand containing products of atomic s-type Slater orbitals and exponentially correlated function of the form $r_{ij} \exp(-\lambda_{ij} r_{ij})$, by employing the Fourier representation of $\exp(-\lambda_{ij} r_{ij})/r_{ij}$ without the use of either the spherical harmonic addition theorem or the Feynman technique. This method is applied to obtain closed-form expressions, in a simple manner, for certain other 2-, 3- and 4-electron atomic integrals with integrands which are products of exponentially correlated functions and atomic s-type Slater orbitals.

## I. INTRODUCTION

One of the most important tasks in quantum theory of many particle atoms is to carry out calculations taking into account the correlation between various electrons in an atom [1]. One novel approach towards the solution of the correlation problem is the use of correlated wave functions which depend explicitly on the interelectronic distance $r_{ij}$. Such a function was employed for the first time in his pioneering work by Hylleraas [2], who calculated energy and wave function for the ground state of helium atom. Over the years this method of approach employing Hylleraas coordinates [3] has been widely applied with success to problems relating to 2-electron systems including hydrogen molecule and helium-like ions with increasing degree of accuracy [4-6]. Evaluation of such 2-electron integrals has been exhaustively discussed by Harris et al. [7].

It seems to be fairly difficult, however, to generalize the method of approach described above, to systems with more than 2 electrons. One of the main problems is that the integrands of the integrals involved in such cases contain several interelectronic distances, making the evaluation of the integrals difficult. The first attempt of application of this correlated wave function method to 3-electron atoms was due to James and Coolidge [8] who investigated the problem relating to the ground state of Li atom. Such a variational calculation relating to the ground state of Be atom was reported by Sims and Hagstrom [9-10] who also discussed the evaluation of certain 3- and 4-electron atomic integrals involving correlation. Several reports relating to evaluation of such 3-and 4-electron atomic integrals involving exponential correlation are available in the literature [11-14]. The atomic 3-electron generating integral containing all three interelectronic distances has been evaluated analytically [15-18], first by Fromm and Hill [15], and later by Remiddi[16], followed by Harris [17]. Also several numerical evaluations of such integrals with increasing degree of accuracy have been reported in the literature [19-24].

In this report, a method is outlined for analytic evaluation of exponentially correlated basic 2-electron integral in a simple manner. The method is also used for evaluating certain other type of 2-, 3- and 4-electron atomic integrals involving exponential correlation. In the process of application of this method for evaluating such integrals, we reproduced the values of some integrals evaluated by others in a different manner.

## II. DEFINITION OF 2-ELECTRON ATOMIC INTEGRAL AND ITS EVALUATION

The general 2-electron integral with integrand containing the product of s-type Slater orbitals and exponentially correlated function is defined as

$$I(i,j,k; \lambda_1, \lambda_2, \lambda_{12}) = \int d\mathbf{r}_1 \int d\mathbf{r}_2\ r_1^{i-1}\ r_2^{j-1}\ r_{12}^{k-1}\ \exp(-\lambda_1 r_1 - \lambda_2 r_2 - \lambda_{12} r_{12})\qquad [1]$$

with integers $i,j,k \geq 0$. Here $r_{12} = |\mathbf{r}_1 - \mathbf{r}_2|$, and $\mathbf{r}_1$ and $\mathbf{r}_2$, respectively, are the position vectors of the two electrons 1 and 2 with respect to the nucleus assumed to be at rest at the origin of the coordinate system. The parameters $\lambda_1, \lambda_2, \lambda_{12}$ should be such that $\lambda_1 + \lambda_2$, $\lambda_1 + \lambda_{12}$ and $\lambda_2 + \lambda_{12}$ are positive[7] although there are no such restrictions on $\lambda_1, \lambda_2, \lambda_{12}$. The simplest integral of this type $I(0,0,0; \lambda_1, \lambda_2, \lambda_{12})$, termed as the basic integral, is given by

$$I(0,0,0; \lambda_1, \lambda_2, \lambda_{12}) = \int d\mathbf{r}_1 \int d\mathbf{r}_2\, (r_1\, r_2\, r_{12})^{-1} \exp(-\lambda_1 r_1 - \lambda_2 r_2 - \lambda_{12} r_{12}) \,. \qquad [2]$$

It has been evaluated by Calais and Lowdin [25] by employing perimetric coordinates, first introduced by Coolidge and James [26], and has the value

$$I(0,0,0; \lambda_1, \lambda_2, \lambda_{12}) = 16\pi^2 [(\lambda_1 + \lambda_2)(\lambda_1 + \lambda_{12})(\lambda_2 + \lambda_{12})]^{-1}. \qquad [3]$$

The value of the general integral in Eq.[1] can be obtained from the value of the basic integral in Eq.[3] through parametric differentiations :

$$I(i,j,k; \lambda_1, \lambda_2, \lambda_{12}) = (-\partial/\partial\lambda_1)^i (-\partial/\partial\lambda_2)^j (-\partial/\partial\lambda_{12})^k\, I(0,0,0; \lambda_1, \lambda_2, \lambda_{12}) \,. \qquad [4]$$

The following particular values obtained from Eq[4] are given here only for the purpose of record:

$$I(0,0,1; \lambda_1, \lambda_2, \lambda_{12}) = 16\pi^2 (\lambda_1 + \lambda_2 + 2\lambda_{12}) [(\lambda_1 + \lambda_2)(\lambda_1 + \lambda_{12})^2 (\lambda_2 + \lambda_{12})^2]^{-1}, \qquad [5]$$

$$I(1,1,1; \lambda_1, \lambda_2, \lambda_{12}) = 64\pi^2 [\lambda_1 \lambda_2 \lambda_{12} + (\lambda_1 + \lambda_2 + \lambda_{12})^3][(\lambda_1 + \lambda_2)^3 (\lambda_1 + \lambda_{12})^3 (\lambda_2 + \lambda_{12})^3]^{-1}, \qquad [6]$$

$$I(1,1,0; \lambda_1, \lambda_2, \lambda_{12}) = 32\pi^2 [\lambda_1 \lambda_2 + (\lambda_1 + \lambda_2 + \lambda_{12})^2][(\lambda_1 + \lambda_2)^3 (\lambda_1 + \lambda_{12})^2 (\lambda_2 + \lambda_{12})^2]^{-1}. \qquad [7]$$

It is observed that the right hand side expressions in Eq.[5-7] are symmetric with interchange of $\lambda_1$ and $\lambda_2$, as expected.

### III. THE PRESENT METHOD OF EVALUATION

In what follows we will outline an alternative simple method of establishing Eq.[3] with the intention of extending the method for obtaining closed-form expressions for certain type of multi-electron atomic integrals involving exponentially correlated functions of $r_{ij}$.

Let us start with Eq.[2] which can be recast as

$$I(0,0,0; \lambda_1, \lambda_2, \lambda_{12}) = \int d\mathbf{r}_1\, r_1^{-1} \exp(-\lambda_1 r_1)\, J(\lambda_2, \lambda_{12}, r_1), \qquad [8]$$

where

$$J(\lambda_2, \lambda_{12}, r_1) = \int d\mathbf{r}_2\, r_2^{-1}\, r_{12}^{-1} \exp(-\lambda_2 r_2 - \lambda_{12} r_{12}). \qquad [9]$$

To obtain a closed-form expression for the integral J above, the following Fourier representation

$$\exp(-\lambda_{12} r_{12})/r_{12} = (1/2\pi^2)\int d\mathbf{k}\, \exp[i\mathbf{k}\cdot(\mathbf{r}_1-\mathbf{r}_2)]/(k^2+\lambda_{12}^2), \qquad [10]$$

where **k** is the Fourier transform variable, is used in Eq.[9]. The orders of integration are then interchanged to evaluate, first the integral over $\mathbf{r}_2$. Subsequently, making use of inverse Fourier transform, the integral over the Fourier transform variable **k** is carried out to obtain

$$J(\lambda_2, \lambda_{12}, r_1) = [4\pi/(\lambda_2^2-\lambda_{12}^2)][\exp(-\lambda_{12}r_1) - \exp(-\lambda_2 r_1)]/r_1. \qquad [11]$$

Substituting Eq.[11] in Eq.[8], the integration over $\mathbf{r}_1$ is performed employing spherical polar coordinates to establish Eq.[3]

## IV. EVALUATION OF SOME OTHER 2-ELECTRON ATOMIC INTEGRALS

(A) It is straightforward to show, using Eqs. [1] and [4] that

$$I(i,j,k;\lambda_1,\lambda_2,\lambda_{12}) = (-\partial/\partial\lambda_1)^i(-\partial/\partial\lambda_2)^j[16\pi^2 k!/(\lambda_2^2-\lambda_1^2)][(\lambda_1+\lambda_{12})^{-k-1} - (\lambda_2+\lambda_{12})^{-k-1}]. \qquad [12]$$

Letting $\lambda_{12} \to 0$ on both sides of Eq.[12], an expression for $I(i,j,k;\lambda_1,\lambda_2,0)$ is obtained which is exactly identical with Eq.(14) of the report of Roberts [13], derived in a different approach, by employing spherical harmonic addition theorem.

(B) A nonsingular integral $I(i,j,-1;\lambda_1,\lambda_2,\lambda_{12})$ of interest, which does not come under general category of Eq.[1], can be evaluated by the method of integration with respect to a parameter. We observe that

$$(-\partial/\partial\lambda_{12})\, I(i,j,-1;\lambda_1,\lambda_2,\lambda_{12}) = (-\partial/\partial\lambda_1)^i(-\partial/\partial\lambda_2)^j\, I(0,0,0;\lambda_1,\lambda_2,\lambda_{12}). \qquad [13]$$

Integrating both sides of Eq.[13] with respect to $\lambda_{12}$ between limits $\lambda_{12}$ and $\infty$, and noting that $I(i,j,-1;\lambda_1,\lambda_2,\lambda_{12})$ goes to zero as $\lambda_{12} \to \infty$, we obtain

$$I(i,j,-1;\lambda_1,\lambda_2,\lambda_{12}) = (-\partial/\partial\lambda_1)^i(-\partial/\partial\lambda_2)^j[16\pi^2/(\lambda_2^2-\lambda_1^2)]\ln[(\lambda_2+\lambda_{12})/(\lambda_1+\lambda_{12})]. \qquad [14]$$

With $i = j = 0$, as a special case, Eq.[14] reduces to

$$I(0,0,-1;\lambda_1,\lambda_2,\lambda_{12}) = [16\pi^2/(\lambda_2^2-\lambda_1^2)]\ln[(\lambda_2+\lambda_{12})/(\lambda_1+\lambda_{12})], \qquad [15]$$

which is in conformity with Eq.(9) of the report of Puchalski and Pachucki [24]. Letting $\lambda_{12} \to 0$ on both sides of Eq.[14], we obtain

$$I(i,j,-1;\lambda_1,\lambda_2,0) = (-\partial/\partial\lambda_1)^i(-\partial/\partial\lambda_2)^j [16\pi^2/(\lambda_2^2-\lambda_1^2)]\ln(\lambda_2/\lambda_1) \qquad [16]$$

which is identical with Eq.(15) of the paper of Roberts [13]. It is worth mentioning here that the integral $I(i,j,-1; \lambda_1, \lambda_2, 0)$ has been evaluated by us directly by using Hylleraas coordinates [2,3] to establish Eq.[16]

(C) There are three other nonsingular integrals of interest, which do not belong to the category of Eq.[1]. These are $I(0,-1,0; \lambda_1, \lambda_2, \lambda_{12})$, $I(-1,0,0; \lambda_1, \lambda_2, \lambda_{12})$ and $I(-1,-1,0; \lambda_1, \lambda_2, \lambda_{12})$.

To evaluate $I(0,-1,0; \lambda_1, \lambda_2, \lambda_{12})$ it is recast as

$$I(0,-1,0; \lambda_1, \lambda_2, \lambda_{12}) = \int d\mathbf{r}_2\, r_2^{-2} \exp(-\lambda_2 r_2)\, J(\lambda_1, \lambda_{12}, r_2), \qquad [17]$$

where J is given by Eqs.[9] and [11]. Substituting the proper closed-form expression for J in Eq.[17], and making use of the standard integral

$$\int_0^\infty (dx/x)[\exp(-ax) - \exp(-bx)] = \ln(b/a), \qquad [18]$$

we obtain

$$I(0,-1,0; \lambda_1, \lambda_2, \lambda_{12}) = [16\pi^2/(\lambda_1^2 - \lambda_{12}^2)]\ln[(\lambda_1+\lambda_2)/(\lambda_2+\lambda_{12})]. \qquad [19]$$

Proceeding in a similar manner we can get

$$I(-1,0,0; \lambda_1, \lambda_2, \lambda_{12}) = [16\pi^2/(\lambda_2^2 - \lambda_{12}^2)]\ln[(\lambda_1+\lambda_2)/(\lambda_1+\lambda_{12})], \qquad [20]$$

which is in conformity with Eq.(12) of the paper of Harris et al. [7].

The value of the integral $I(-1,-1,0; \lambda_1, \lambda_2, \lambda_{12})$ has already been reported by Fromm and Hill [15] and by Harris et al.[7].

## V. EVALUATION OF SOME 3-ELECTRON ATOMIC INTEGRALS

We now consider evaluation of a nine-dimensional integral over the variables $\mathbf{r}_1$, $\mathbf{r}_2$ and $\mathbf{r}_3$ with an integrand involving exponential correlation relating to only two of the three interelectronic distances. For example we take as basic integral

$$K(\lambda_1,\lambda_2,\lambda_3; \lambda_{12},\lambda_{23}) = \int d\mathbf{r}_1 \int d\mathbf{r}_2 \int d\mathbf{r}_3 (r_1\, r_2\, r_3\, r_{12}\, r_{23})^{-1}$$
$$\times \exp(-\lambda_1 r_1 -\lambda_2 r_2 -\lambda_3 r_3 -\lambda_{12} r_{12} -\lambda_{23} r_{23}). \qquad [21]$$

The integral K can be recast as

$$K = \int d\mathbf{r}_2 (r_2)^{-1} \exp(-\lambda_2 r_2)\, J(\lambda_1, \lambda_{12}, r_2)\, J(\lambda_3, \lambda_{23}, r_2), \qquad [22]$$

where J is given by Eqs. [9] and [11]. Making proper substitution for J's in Eq.[22], it just reduces to a three - dimensional integral over the variable $\mathbf{r}_2$. Performing the angular integration easily and then using the standard result given by Eq.[18], the following closed-form expression for K is obtained :

$$K = 64\pi^3 [(\lambda_1^2 - \lambda_{12}^2)(\lambda_3^2 - \lambda_{23}^2)]^{-1}$$

$$\times \ln[(\lambda_2 + \lambda_3 + \lambda_{12})(\lambda_1 + \lambda_2 + \lambda_{23})(\lambda_2 + \lambda_{12} + \lambda_{23})^{-1}(\lambda_1 + \lambda_2 + \lambda_3)^{-1}]. \qquad [23]$$

Differentiating both sides of Eq.[23] with respect to (- $\lambda_2$), an expression for the integral (-$\partial K/\partial \lambda_2$) is easily obtained which is identical with the right hand side of Eq.(9) of report of Bonham [14], discarding all the parametric differentiations therein. This proves that the derivation of Eq.[23] is correct.

We would like to point out here that, if only one or two of the three interelectronic distances are involved in the integrand, the integral can be evaluated easily as outlined here. However, analytic evaluation of the basic 3-electron atomic integral with exponential correlation involving all the three interelectronic distances has already been done and reported in the literature [15-18] though the algebra involved is quite complicated.

## VI. EVALUATION OF SOME 4-ELECTRON ATOMIC INTEGRALS

(A) Let us consider evaluation of a 4-electron atomic integral L given by

$$L = \int d\mathbf{r}_1 \int d\mathbf{r}_2 \int d\mathbf{r}_3 \int d\mathbf{r}_4 \, (r_1 r_2 r_3 r_4 r_{12} r_{23} r_{24})^{-1} \exp(-\lambda_1 r_1 - \lambda_2 r_2 - \lambda_3 r_3 - \lambda_4 r_4 - \lambda_{12} r_{12} - \lambda_{23} r_{23} - \lambda_{24} r_{24}), \qquad [24]$$

which can be recast as

$$L = \int d\mathbf{r}_2 (r_2)^{-1} \exp(-\lambda_2 r_2) \, J(\lambda_1, \lambda_{12}, r_2) \, J(\lambda_3, \lambda_{23}, r_2) \, J(\lambda_4, \lambda_{24}, r_2), \qquad [25]$$

with J's given by Eqs.[9] and [11], as before. After proper substitutions for J's in Eq.[25] the angular integration is done to reduce the integral L to one-dimensional form

$$L = A \int_0^\infty dr_2 \, r_2^{-2} f(r_2), \qquad [26]$$

where $A = 256\pi^4 [(\lambda_1^2 - \lambda_{12}^2)(\lambda_3^2 - \lambda_{23}^2)(\lambda_4^2 - \lambda_{24}^2)]^{-1}$, [27]

and $f(r_2)$ is a sum of eight terms of the form $\exp(-\beta_i r_2)$, i = 1,2,3, …….8, with $\beta_i$ being a sum of four different $\lambda$'s out of the seven $\lambda$'s in Eq.[24]. Also all $\beta_i$ values are different.

It can be shown that as $r_2 \to 0$, all the three functions $f(r_2)$, $f_1(r_2)$ and $f_2(r_2) \to 0$. Here $f_1$ and $f_2$ represent 1$^{st}$ order and 2$^{nd}$ order differentiation, respectively. Also by employing L' Hospital's rule for 0/0, it can be proved that $f(r_2)/r_2^2 = 0$ and $f(r_2)/r_2 = 0$ as $r_2 \to 0$. Integrating by parts, and then making use of Eq.[18], the following closed-form expression for the integral L is obtained :

$$L = A \Sigma_i L_i \qquad [28]$$

where A is given by Eq.[27], and

$L_1 = (\lambda_2+\lambda_{12}+\lambda_{23}) \ln [(\lambda_2+\lambda_{12}+\lambda_{23}+\lambda_{24})/ (\lambda_2+\lambda_{12}+\lambda_{23}+\lambda_4)]$,

$L_2 = (\lambda_2+\lambda_{12}+\lambda_3) \ln [(\lambda_2+\lambda_{12}+\lambda_3+\lambda_4)/ (\lambda_2+\lambda_{12}+\lambda_3+\lambda_{24})]$,

$L_3 = (\lambda_2+\lambda_1+\lambda_{23}) \ln [(\lambda_2+\lambda_1+\lambda_{23}+\lambda_4)/ (\lambda_2+\lambda_1+\lambda_{23}+\lambda_{24})]$,

$L_4 = (\lambda_1+\lambda_2+\lambda_3) \ln [(\lambda_1+\lambda_2+\lambda_3+\lambda_{24})/ (\lambda_1+\lambda_2+\lambda_3+\lambda_4)]$,

$L_5 = \lambda_{24} \ln [(\lambda_2+\lambda_{12}+\lambda_{23}+\lambda_{24})/ (\lambda_2+\lambda_{12}+\lambda_3+\lambda_{24})]$,

$L_6 = \lambda_{24} \ln [(\lambda_1+\lambda_2+\lambda_3+\lambda_{24})/ (\lambda_1+\lambda_2+\lambda_{23}+\lambda_{24})]$,

$L_7 = \lambda_4 \ln [(\lambda_2+\lambda_{12}+\lambda_3+\lambda_4)/ (\lambda_2+\lambda_{12}+\lambda_{23}+\lambda_4)]$,

$L_8 = \lambda_4 \ln [(\lambda_2+\lambda_1+\lambda_{23}+\lambda_4)/ (\lambda_1+\lambda_2+\lambda_3+\lambda_4)]$.

Differentiating both sides of Eq.[28] with respect to $(-\lambda_2)$, we get a closed-form expression for the integral $(-\partial L/\partial \lambda_2)$. Then putting $\lambda_{12} = \lambda_{23} = \lambda_{24} = 0$, as a special case, on both sides of the equation containing the integral $(-\partial L/\partial \lambda_2)$ and its closed-form expression, we are able to reproduce an expression in conformity with Eq.(16) of the paper by Roberts [13]. This proves that the value of the integral L obtained by us is correct.

(B) Let us evaluate another 4-electron atomic integral M given by

$$M = \int d\mathbf{r}_1 \int d\mathbf{r}_2 \int d\mathbf{r}_3 \int d\mathbf{r}_4 \, (r_1 r_2 r_4 r_{12} r_{23} r_{34})^{-1} \exp(-\lambda_1 r_1 - \lambda_2 r_2 - \lambda_3 r_3 - \lambda_4 r_4 - \lambda_{12} r_{12} - \lambda_{23} r_{23} - \lambda_{34} r_{34}). \qquad [29]$$

The integral M can be recast as

$$M = \int d\mathbf{r}_2 (r_2)^{-1} \exp(-\lambda_2 r_2) \, J(\lambda_1, \lambda_{12}, r_2) \, N(\lambda_3, \lambda_{23}, \lambda_4, \lambda_{34}, r_2), \qquad [30]$$

where

$$N = \int d\mathbf{r}_3 (r_{23})^{-1} \exp(-\lambda_3 r_3 - \lambda_{23} r_{23}) \, J(\lambda_4, \lambda_{34}, r_3), \qquad [31]$$

with J given by Eqs.[9] and [11]. A closed-form expression for the integral N is easily obtained which comes out as a function of $r_2$. Substituting this expression for N in Eq.[30], the angular integration is performed to reduce the integral M to one-dimensional form, which is subsequently evaluated by the use of Eq.[18]. We obtained the following closed-form expression for the integral M:

$$M = 256\pi^4[(\lambda_1^2 - \lambda_{12}^2)(\lambda_4^2 - \lambda_{34}^2)]^{-1}[M_1 - M_2] \qquad [32]$$

where

$$M_1 = [(\lambda_3+\lambda_{34})^2 - \lambda_{23}^2]^{-1} \ln[(\lambda_1+\lambda_2+\lambda_{23})(\lambda_2+\lambda_3+\lambda_{12}+\lambda_{34})(\lambda_2+\lambda_{12}+\lambda_{23})^{-1}(\lambda_1+\lambda_2+\lambda_3+\lambda_{34})^{-1}],$$

and

$$M_2 = [(\lambda_3+\lambda_4)^2 - \lambda_{23}^2]^{-1} \ln[(\lambda_1+\lambda_2+\lambda_{23})(\lambda_2+\lambda_3+\lambda_4+\lambda_{12})(\lambda_2+\lambda_{12}+\lambda_{23})^{-1}(\lambda_1+\lambda_2+\lambda_3+\lambda_4)^{-1}].$$

Differentiating both sides of Eq.[32] with respect to $(-\lambda_2)$, we obtain a closed-form expression for another 4-electron integral $(-\partial M/\partial \lambda_2)$ in conformity with Eq,(11) of paper of Bonham[14]. This observation establishes the validity of the derivation of Eq.[32].

## VII. CONCLUSIONS

Several 2-,3- and 4-electron atomic integrals involving exponential correlation and products of s-type Slater orbitals have been evaluated in a simple manner without expanding $r_{ij}$'s or its powers in terms of spherical harmonics, unlike many published papers, where spherical harmonic addition theorem has been used. Also we did not use Feynman technique [27]. Reduction of these integrals to closed-form in the present work is consistent with the conjecture 'A' put forward by Bonham[14]. The method of evaluation presented in this report can be extended to some other multi-electron atomic integrals involving s-type Slater orbitals and exponential correlation. We also find that the method can be easily extended to evaluation of certain 2-, 3-, and 4-electron atomic integrals involving exponential correlation and products of p-, d-, f-type Slater orbitals.

### ACKNOWLEDGEMENT

The author is extremely thankful to Prof. D. K. Rai, under whose supervision he obtained his Ph.D., and, who has been a constant source of encouragement till date.